# REFINED CHEMICAL ABUNDANCES OF π DRA AND HR 7545 WITH ATLAS12

*\*Doğuş Özuyar, \*Aslı Elmaslı, \*Şeyma Çalışkan*
*\*Department of Astronomy and Space Sciences,*
*Ankara University, Science Faculty*
*Ankara, Turkey*

## ABSTRACT

*We present the more refined abundances of elements (C, O, Na, Mg, Al, Si, Ca, Sc, Ti, V, Cr, Mn, Fe, Ni, Zn, Sr, Y, Zr and Ba) determined using the computationally intensive ATLAS12 atmosphere model for the metallic-line stars π Dra and HR 7545. We compare the chemical abundances derived from ATLAS9 model atmosphere with those of derived from ATLAS12. The abundances are in agreement with each other, within their uncertainties. We thus state that ATLAS9 may be used for the abundance analysis of slowly rotating chemically peculiar stars, such as π Dra and HR 7545.*

## INTRODUCTION

One of A-type star sub-groups, called metallic-line A-type (hereafter Am) stars, exhibits generally A-type spectral behavior with some elemental anomalies. These anomalies are typically deficiency of Sc and sometimes Ca, under-abundances in the light elements (C, O, etc.), followed by an increase in iron-peak element (Cr, Mn, Fe, Ni, etc.) abundances, and a considerably overabundances in the heavy elements, such as Sr, Y, Zr, and Ba, with respect to the solar values [1]. Another characteristic property of Am stars is their low rotational velocities compared to normal A-type stars ($< 120$ km s$^{-1}$, [2]). Also, the Am stars are generally found in close binary systems.

π Dra, which has been rarely studied for the past 30 years, was classified as an A2mA3 IV-type star by [3]. They reported that the star could be a notable object due to its relatively sharp spectral lines among the rapidly rotating early A-type stars. Reference [4] carried out an abundance analysis of π Dra, covering the wavelength from 3815 to 4590 Å. They stated that the star showed an excessive amount of heavy metals and therefore it was an Am-type star. Later, reference [5] derived element abundances of π Dra, analyzing the wavelengths between 3830 and 4740 Å. He confirmed the great heavy-element anomalies seen in the star. Subsequent to this study, reference [6] expanded the wavelength range up to 4937 Å for a spectroscopic analysis, and found a good agreement between the results of the extended and his previous studies. Recently, in the detailed abundance analysis, reference [7] analyzed a much wider wavelength range from 3900 to 9100 Å, obtained at the TÜBİTAK National Observatory (TUG). They detected an under-abundance in Sc relative to the Sun in addition to a notable overabundance in iron peak (V, Cr, Mn, Ni) and heavy elements (Zn, Y, Zr, Sr, Ba). All of the analyses above ([4], [5], [6], and [7]) have revealed that π Dra is a chemically peculiar Am star.

HR 7545 was ascribed to a spectral type A2 III by [8]. Its atmospheric parameters were determined in the work [9] as effective temperature $T_{eff}$ of 8579 K, surface gravity log g of 3.25 and [Fe/H] of -0.05. Reference [10] measured the $T_{eff}$ as 8585 K. Rotational velocity of HR 7545 was given by [11] as 15 kms$^{-1}$ and the radial velocity was reported as -4.0kms$^{-1}$ [12]. The first detailed abundance analysis of HR 7545 carried out by [7], revealed the star's metallic-line nature from the high resolution spectroscopic data.

In this study, we aim to derive more refined abundances for two Am stars, π Dra and HR 7545, with ATLAS12 model atmosphere code.

## MODEL ATMOSPHERES

ATLAS9 is a model atmosphere code for computing models with the opacity distribution function (ODF) method. ATLAS9 model atmospheres assume a plane parallel geometry, hydrostatic equilibrium and line formation in local thermodynamic equilibrium. The models include the line opacity tables in the form of opacity distribution functions based on the solar abundances. The advantage of this code is that its computation duration relatively takes shorter time. In other respect, one cannot compute a model with arbitrary abundances using ATLAS9 [13]. In the other word, ATLAS9 does not allow to compute a model having different chemical composition profile from the Sun. However, the abundances of chemically peculiar stars are quite different from those of the Sun.

On the other hand, ATLAS12 [14] uses the opacity sampling (OS) method for computing models. It considers a plane parallel geometry, hydrostatic equilibrium and line formation in local thermodynamic equilibrium similar to ATLAS9. An opacity sampling model atmosphere program ATLAS12 allows computation of models with individual abundances using line data. Briefly, the code allows arbitrary abundances. This main difference (between ATLAS9 and ATLAS12) makes ATLAS12 reliable in computing model atmosphere for chemically peculiar stars.

For these reasons, we use ATLAS12 to compute model atmospheres having individual abundances of our chemically peculiar stars π Dra and HR 7545.

## CHEMICAL ABUNDANCE ANALYSIS

For the analysis of both Am stars, we used the line list of [7]. The atmospheric parameters (effective temperature, surface gravity, microturbulent velocity, and metallicity) and rotational velocities (see Table I) as well as the measurements of equivalent width (EW) of the unblended absorption lines in the spectra of the stars were adopted from the same study.

Table I. The atmospheric parameters and rotational velocities of our targets [7].

| Star | $T_{eff}$ (K) | $\log g$ (dex) | $\xi$ (kms$^{-1}$) | [Fe/H] | $v\sin i$ (kms$^{-1}$) |
|---|---|---|---|---|---|
| π Dra | 9200 | 3.9 | 3.1 | 0.4 | 25±1 |
| HR 7545 | 8800 | 3.4 | 3.4 | 0.2 | 12 |

We used WIDTH9 code [14] to derive the abundances from EW measurements of individual lines for each star. We adopted an uncertainty of 0.15 dex for species that also with only one detected line. Using available ATLAS9 models and abundances of the stars, we produced the atmosphere models with ATLAS12 code for π Dra and HR 7545. Then, we re-determined the abundances using ATLAS12 models, in order to achieve more refined abundance values of two Am stars. The available abundances from ATLAS9 and the abundances derived from ATLAS12 are listed in Table II with the standard deviations.

Table II. Derived element abundances from ATLAS9 and ATLAS12 for π Dra. N is number of lines.

| Atomic Species | π Dra | | | HR 7545 | | |
|---|---|---|---|---|---|---|
| | $\log \varepsilon_{ATLAS9}$ | $\log \varepsilon_{ATLAS12}$ | N | $\log \varepsilon_{ATLAS9}$ | $\log \varepsilon_{ATLAS12}$ | N |
| C I | -- | -- | -- | 8.20±0.09 | 8.24±0.09 | 3 |
| O I | -- | -- | -- | 8.56±0.10 | 8.56±0.10 | 2 |
| Na I | 7.51±0.11 | 7.53±0.11 | 2 | 6.70±0.15 | 6.75±0.15 | 1 |
| Mg I | 7.70±0.09 | 7.79±0.08 | 4 | 7.86±0.15 | 7.85±0.15 | 1 |
| Al II | -- | -- | -- | 6.53±0.15 | 6.51±0.15 | 1 |
| Si II | 7.91±0.14 | 7.89±0.14 | 3 | 7.39±0.04 | 7.38±0.04 | 3 |
| Ca I | 6.64±0.11 | 6.78±0.11 | 5 | 6.43±0.03 | 6.49±0.03 | 9 |
| Ca II | -- | -- | -- | 6.44±0.13 | 6.46±0.13 | 5 |
| Sc II | 2.57±0.12 | 2.66±0.12 | 2 | 2.99±0.14 | 3.02±0.13 | 5 |
| Ti II | 5.28±0.16 | 5.34±0.16 | 12 | 5.04±0.13 | 5.07±0.13 | 26 |
| V II | -- | -- | -- | 4.27±0.03 | 4.30±0.03 | 4 |
| Cr I | 6.16±0.10 | 6.28±0.15 | 1 | 5.66±0.08 | 5.72±0.08 | 4 |
| Cr II | 6.16±0.10 | 6.20±0.18 | 17 | 5.70±0.10 | 5.72±0.10 | 8 |
| Mn I | -- | -- | -- | 5.47±0.12 | 5.52±0.12 | 3 |
| Mn II | -- | -- | -- | 5.59±0.15 | 5.60±0.15 | 1 |
| Fe I | 7.94±0.17 | 8.05±0.17 | 37 | 7.71±0.13 | 7.76±0.13 | 119 |
| Fe II | 7.93±0.18 | 7.96±0.18 | 47 | 7.69±0.11 | 7.70±0.11 | 60 |
| Ni I | 7.01±0.05 | 7.13±0.05 | 3 | 6.73±0.12 | 6.78±0.12 | 11 |
| Ni II | -- | -- | -- | 6.77±0.15 | 6.79±0.15 | 1 |
| Zn I | -- | -- | -- | 5.21±0.03 | 5.26±0.03 | 2 |
| Sr II | 4.24±0.10 | 4.32±0.15 | 1 | 3.75±0.15 | 3.80±0.15 | 1 |
| Y II | 3.51±0.10 | 3.59±0.15 | 1 | 3.13±0.15 | 3.17±0.15 | 7 |
| Zr II | -- | -- | -- | 3.42±0.06 | 3.45±0.06 | 2 |
| Ba II | 3.71±0.20 | 3.78±0.19 | 2 | 3.67±0.09 | 3.72±0.10 | 2 |

## RESULTS AND CONCLUSION

The abundance pattern of both π Dra and HR 7545 showed deviations from the solar abundances[15], as expected Am stars. The Scandium abundance ([Sc/H] = -0.52) was under-abundant compared to the solar photospheric value. The iron-peak element abundances (Cr, Fe, Ni) were larger than solar values. Its abundance pattern also showed substantially overabundances of Sr, Y, and Ba with respect to the solar abundances. The chemical abundances of HR 7545 exhibited an underabundance in C and O, as well as an increase in Na and some iron-peak elements V, Ni, and Zn. [Sc/H] and [Ca/H] values of the star were -0.15 and 0.10. The Magnesium, Manganese, and Iron abundances of the star showed slightly overabundance with ~0.2. The Al, Ti, and Cr abundances were the solar. The star had a significant overabundance in the heavy elements Sr, Y, Zr, and Ba. All these abundance values of π Dra and HR 7545 were not different from those of derived with ATLAS9. As shown in Fig. 1, the differences between the abundances of ATLAS9 and ATLAS12 .

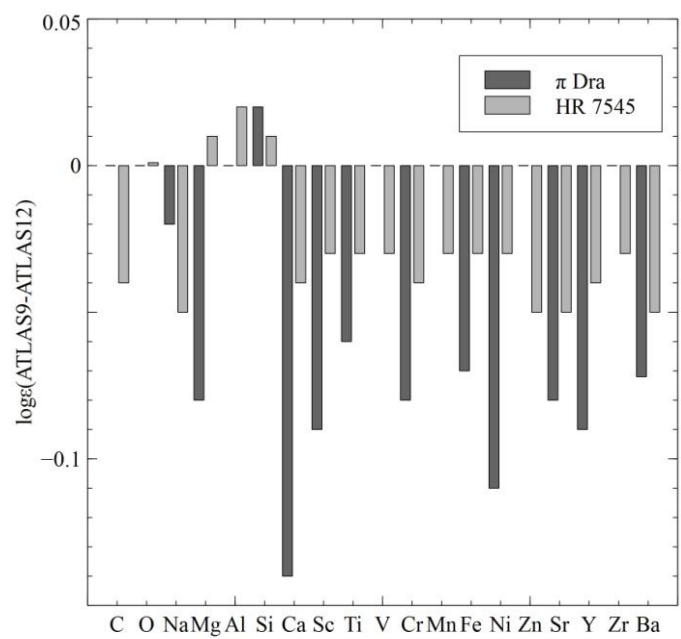

Figure 1.  Differences between abundances from ATLAS9 and ATLAS12.

As a results, we derived the abundances of two Am-type stars π Dra and HR7545 from ATLAS12 model atmospheres using their EW measurements. The abundances of individual elements derived from ATLAS9 and ATLAS12 are compatible with each other within their uncertainties (< 0.15 dex) for each star. We thus confirm the result of [16] that ATLAS9 may be used to measure the abundances for moderate chemically peculiar stars. More abundance studies of new chemical peculiar stars, such as Am-type, will allow us to understand the origin of their interesting abundance patterns.

## ACKNOWLEDGMENT

The authors thank TÜBİTAK National Observatory for a partial support in using RTT150 (Russian-Turkish 1.5-m telescope in Antalya) with project number 14BRTT150-671.